\documentclass[english,12pt]{iopart}
\usepackage[T1]{fontenc}
\usepackage[latin1]{inputenc}
\usepackage{longtable}
\usepackage{graphicx}

\makeatletter

\providecommand{\tabularnewline}{\\}

\usepackage{babel}
\makeatother
\begin{document}

\title{A first-principles comparison of the electronic properties of $MgC_{y}Ni_{3}$
and $ZnC_{y}Ni_{3}$ alloys }

\author{P. Jiji Thomas Joseph and Prabhakar P. Singh}

\address{Department of Physics, Indian Institute of Technology Bombay, Mumbai
India 400076}

\begin{abstract}
First-principles, density-functional-based electronic structure calculations
are employed to study the changes in the electronic properties of
$ZnC_{y}Ni_{3}$ and $MgC_{y}Ni_{3}$ using the Korringa-Kohn-Rostoker
coherent-potential approximation method in the atomic sphere approximation
(KKR-ASA CPA). As a function of decreasing $C$ $at$\%, we find a
steady decrease in the lattice constant and bulk modulus in either
alloys. However, the pressure derivative of the bulk modulus displays
an opposite trend. Following the Debye model, which relates the pressure
derivative of the bulk modulus with the average phonon frequency of
the crystal, it can thus be argued that $ZnCNi_{3}$ and its disordered
alloys posses a different phonon spectra in comparison to its $MgCNi_{3}$
counterparts. This is further justified by the marked similarity we
find in the electronic structure properties such as the variation
in the density of states and the Hopfield parameters calculated for
these alloys. The effects on the equation of state parameters and
the density of states at the Fermi energy, for partial replacement
of $Mg$ by $Zn$ are also discussed. 
\end{abstract}
\maketitle

\section{Introduction}

In spite of being iso-structural and iso-valent to the cubic perovskite
$8$K superconductor $MgCNi_{3}$ \cite{Nature-411-54}, $ZnCNi_{3}$
remains in the normal metal state down to $2$K \cite{sst-17-274}.
The specific heat measurements indicate that the absence of superconductivity
in $ZnCNi_{3}$ may be due to a substantial decrease in the density
of states at the Fermi energy $N(E_{F})$ resulting from its relatively
low unit cell volume in comparison with $MgCNi_{3}$ \cite{sst-17-274}.
However, electronic structure calculations show that the decrease
in $N(E_{F})$ is not sizable enough to make $ZnCNi_{3}$ non-superconducting
\cite{PRB-70-060507}. For both $MgCNi_{3}$ \cite{PRB-140507,PRL-027001,PRB-100508,JPCM-L595}
and $ZnCNi_{3}$ \cite{PRB-70-060507} the density of states spectra
display similar characteristics, particularly in the distribution
of electronic states near the Fermi energy $E_{F}$. The electronic
states at $E_{F}$ are dominated by $Ni$ $3d$ states with a little
admixture of $C$ $2p$ states. There exists a strong van Hove singularity-like
feature just below $E_{F}$, which is primarily derived from the $Ni$
$3d$ bands. 

To account for the lack of superconductivity in $ZnCNi_{3}$, the
density-functional based calculations emphasize that the material
subjected to the specific heat measurements may be non-stoichiometric
in the $C$ sub-lattice \cite{PRB-70-060507}. This would then make
it similar to the $\alpha-$ phase of $MgCNi_{3}$, which has a low
unit cell volume and remains non- superconducting \cite{PhysicaC-1}.
It has been shown earlier that exact $C$ content in $MgC_{y}Ni_{3}$
depends on the nature of synthesis and other experimental conditions
\cite{Nature-411-54,PhysicaC-1,PRB-052506,PRB-172507,JJAP-L1365,PRB-024523}.
According to Johannes and Pickett \cite{PRB-70-060507}, the arguments
that favor non-stoichiometry are the following: (i) Total energy minimization
en-route to equilibrium lattice constant within the local-density
approximation (LDA) finds an overestimated value for $ZnCNi_{3}$
in comparison with the experimental values. In general, overestimation
is not so common in LDA. Meanwhile, when one uses similar technique
for $MgCNi_{3}$, the calculations find a slightly underestimated
value which is consistent within the limitations of the density-functional
theory \cite{PRB-140507,PRB-72-064519,PRB-72-214206}. (ii) The authors
also find $N(E_{F})$ in $MgCNi_{3}$ estimated as $13.6$ states/Ry
atom, while for $ZnCNi_{3}$, under similar approximations, it was
found to be $11.01$ states/Ry atom. Note that it has been shown both
experimentally as well as from first-principles calculations that
a decrease in the lattice constant or a decrease in the $C$ occupancy
would lead to a decrease in $N(E_{F})$ \cite{PRB-72-064519}. (iii)
A decrease in the unit cell dimensions can induce phonon hardening.
This is well supported by the experiments which find the Debye temperature
approximately 1.6 times higher for $ZnCNi_{3}$ in comparison to $MgCNi_{3}$\cite{sst-17-274}. 

Earlier synthesis of $ZnC_{y}Ni_{3}$ \cite{ZMetallk-50-1959,acta-met-7-415,Metall-15-124}
finds the lattice constant to be $6.899$ a.u., for which the occupancy
in the $C$ sub-lattice was just $70$\%. The authors have employed
similar preparation technique for $MgCNi_{3}$ \cite{ZMetallk-50-1959}
and have found that the $C$ occupancy ranges between $0.5$-$1.25,$
which is consistent with the recent reports \cite{Nature-411-54,PhysicaC-1,PRB-052506,PRB-172507,JJAP-L1365,PRB-024523,SSC-121-73}.
Lattice constant for $ZnCNi_{3}$, as high as $7.126$ a.u. has also
been reported elsewhere \cite{ZMettalk-52-477,LBseries}, which then
becomes consistent with the recent total energy minimized value using
density-functional based methods. Hence, it seems that $ZnCNi_{3}$
which was subjected to specific heat experiments \cite{sst-17-274}
may indeed suffer from non-stoichiometry.

To understand and compare the effects of $C$ stoichiometry on the
structural and electronic properties of $MgC_{y}Ni_{3}$ and $ZnC_{y}Ni_{3}$,
we carry out a detail study using the Korringa-Kohn-Rostoker (KKR)
Green's function method \cite{Physica-13-392,PR-94-1111} formulated
in the atomic sphere approximation (ASA) \cite{kluwer-1997}. For
disorder, we employ the coherent-potential approximation (CPA) \cite{PR-156-809}.
Characterization of $MgC_{y}Ni_{3}$ and $ZnC_{y}Ni_{3}$ with $0.85\leq y\leq1.00$
mainly involves the changes in the equation of state parameters viz.,
the equilibrium lattice constant, bulk modulus and its pressure derivative.
The electronic structure is studied with the help of total and sub-lattice
resolved density of states. The propensity of magnetism in these materials
is studied with the help of fixed-spin moment method \cite{JPF-14-129}
in conjunction with the Landau theory of phase transition \cite{Landau}.
The Hopfield parameter $\eta$ which generally maps the local ``chemical''
property of an atom in a crystal is also calculated as suggested by
Skriver and Mertig \cite{PRB-4431}, and its variation as a function
of lattice constant has also been studied. In general, we find that
both $MgCNi_{3}$ and $ZnCNi_{3}$ display very similar electronic
structure. Evidences point that the non-superconducting nature of
$ZnCNi_{3}$ may be related to the crystal structure characteristics,
namely phonon spectra.

\section{Computational details}

The ground state properties of $MgC_{y}Ni_{3}$ and $ZnC_{y}Ni_{3}$
are calculated using the KKR-ASA-CPA method of alloy theory. For improving
alloy energetics, the ASA is corrected by the use of both the muffin-tin
correction for the Madelung energy \cite{PRL-55-600} and the multi-pole
moment correction to the Madelung potential and energy \cite{PRB-66-024201,PRB-66-024202}.
These corrections have brought significant improvement in the accuracy
of the total energy by taking into account the non-spherical part
of polarization effects \cite{CMC-15-119}. The partial waves in the
KKR-ASA calculations are expanded up to $l_{max}=3$ inside atomic
spheres, although the multi-pole moments of the electron density have
been determined up to $l_{max}^{M}=6,$ which is used for the multi-pole
moment correction to the Madelung energy. In general, the exchange-correlation
effects are taken into consideration via the local-density approximation
with Perdew and Wang parametrization \cite{PRB-45-13244}, although
a comparison in the equation of state parameters has been made in
this work with the generalized gradient approximation (GGA) \cite{prb-54-16533}.
The core states have been recalculated after each iteration. The calculations
are partially scalar-relativistic in the sense that although the wave
functions are non-relativistic, first order perturbation corrections
to the energy eigenvalues due to the Darwin and the mass-velocity
terms are included. The atomic sphere radii of $Mg$ ($Zn$), $C$
and $Ni$ were kept as $1.404$, $0.747$, and $0.849$ of the Wigner-
Seitz radius, respectively. The vacancies in the $C$ sub-lattice
are modeled with the help of empty spheres, and their radius is kept
same as that of $C$ itself. The overlap volume resulting from the
blow up of the atomic spheres was less than $15$\%, which is legitimate
within the accuracy of the approximation \cite{LMTO-1984}. 

The electron-phonon coupling parameter $\lambda$ can be expressed
as $\eta$/$M\left\langle \omega^{2}\right\rangle $, where $\eta$
is the Hopfield parameter, expressed as the product of $N(E_{F})$
and the mean square electron-ion matrix element $\left\langle I^{2}\right\rangle $,
with $M$ and $\left\langle \omega^{2}\right\rangle $ being the ionic
mass and average phonon frequency \cite{douglass}. However, one may
note that the above decomposition of the problem into electronic and
phonon contributions is only approximate since in principle $\left\langle \omega^{2}\right\rangle $
is also determined by the electronic states. It follows that the Hopfield
parameter is the most simple basic quantity which one may obtain from
first-principles as suggested by Gaspari and Gyorffy \cite{PRL-28-801}.
The latter assumes a rigid muffin-tin approximation (RMTA) in which
the potential enclosed by a sphere rigidly moves with the ion and
the change in the crystal potential, caused by the displacement, is
given by the potential gradient. Within the RMTA the spherically averaged
part of the Hopfield parameter may be calculated as, 

\begin{equation}
\eta_{0}=\frac{2}{N(E_{F})}\sum_{l}(l+1)M_{l,l+1}^{2}\frac{N_{l}(E_{F})N_{l+1}(E_{F})}{(2l+1)(2l+3)}\label{eq-123d}\end{equation}
where $N(E_{F})$ is the total density of state per spin at the Fermi
energy, and $N_{l}$ the $l^{th}$ partial density of state calculated
at the Fermi energy $E_{F}$, on the site considered. The term $M_{l,l+1}$
is the electron-phonon matrix element given as \cite{PRB-4431},

\begin{equation}
M_{l,l+1}=\int_{0}^{S}r^{2}R_{l}\frac{dV}{dr}R_{l+1}dr\label{eq-123m}\end{equation}
which are obtained from the gradient of the potential and the radial
solutions $R_{l}$ and $R_{l+1}$ of the Schrodinger equation evaluated
at $E_{F}$. The special form of the Eqs.\ref{eq-123d} and \ref{eq-123m}
stems from the ASA in which the radial wave functions are normalised
to unity in the atomic sphere of radius $S$, i.e., $\int_{0}^{S}r^{2}R_{l}^{2}(r)dr$=$1$.
In ASA, $M_{l,L+1}$ is expressed in terms of logarithmic derivatives
$D_{l}$=$rR_{l}^{'}$/$R_{l}$ evaluated at the sphere boundary.
Skriver and Mertig derive the expression for $M_{l,l+1}$ as

\begin{equation}
\begin{array}{cc}
M_{l,l+1}= & -\phi_{l}(E_{F})\phi_{l+1}(E_{F})\times\\
 & \left\{ \left[D_{l}-l)\right]\left[D_{l+1}+l+2)\right]+\left[E_{F}-V(S)\right]S^{2}\right\} \end{array}\label{eq-234r}\end{equation}
where $V(S)$ is the one-electron potential and $\phi_{l}(E_{F})$
the sphere boundary amplitude of the $l$ partial wave evaluated at
$E_{F}$. 

Numerical estimate to the magnetic energy are carried out using the
fixed-spin-moment method \cite{JPF-14-129}. In the fixed-spin-moment
method the total energy is obtained for a given magnetization $M$,
i.e., by fixing the numbers of electrons with up and down spins. In
this case, the Fermi energies in the up and down spin bands are not
equal to each other because the equilibrium condition would not be
satisfied for arbitrary $M$. At the equilibrium $M$ two Fermi energies
will coincide with each other. The total magnetic energy becomes minimum
or maximum at this value of $M$. Note that the two approaches, i.e.,
the self-consistent, floating-spin-moment method as well as the fixed-spin
moment-method are equivalent in the sense that for a given lattice
constant the magnetic moment calculated by the standard floating-spin
moment approach is the same as the magnetic moment for which the fixed-spin
moment total energy has its minimum \cite{PRB-58-4341}. In practice,
the floating-spin moment approach sometimes runs into some convergence
problem. From experience, to avoid such predicaments in convergence,
one may carefully monitor the mixing of the initial and final charges
during the iterations and increase the number of $\mathbf{k}-$ points.
Thus, for a better resolution to determine the change in the total
energy with respect to the input magnetization, the $\mathbf{k}$-
mesh had $1771$ $\mathbf{k}-$ points in the irreducible wedge of
the cubic Brillouin zone. 

By the fixed-spin-moment method the difference $\Delta E(M)$ (=$E(M)-E(0)$)
for given values of $M$ is calculated. The calculated $\Delta E(M)$
is fitted to the phenomenological Landau equation of phase transition
which is given as

\begin{equation}
\Delta E(M)=\sum_{n>0}\,\,\frac{1}{2n}a_{2n}M^{2n}\label{eq-643f}\end{equation}
for $n=3$, where the sign of the coefficient $a_{2n}$ for $n=1$
determines the nature of the magnetic ground state, i.e., $a_{2}$>0
refers to a paramagnetic ground state while $a_{2}<0$ refers to a
ferromagnetic phase. We have applied the approach described above
to the study of carbon vacancy in $MgCNi_{3}$ \cite{PRB-72-064519}
and $3d$ transition-metal-$MgCNi_{3}$ alloys \cite{PRB-72-214206}.

\section{Results and Discussion}

\subsection{Equation of state}

Both X-ray and neutron diffraction techniques unambiguously report
$MgCNi_{3}$ and $ZnCNi_{3}$ as cubic perovskites with their lattice
constants determined as $7.201$ and $6.918$ a.u., respectively.
Assuming an underlying rigid cubic lattice, with $Mg$($Zn$) at cube
corners, $Ni$ at the faces and $C$ at the octahedral interstitial
site, the total energy minimization were carried out to determine
the equation of state parameters. The total energies calculated, self-consistently,
for six lattice constants close to equilibrium were fed as input to
a third-order Birch-Murnaghan equation of state \cite{JGR-57-227,FDES-140}.
Note that the Birch-Murnaghan equation is derived from the theory
of finite strain, by considering an elastic isotropic medium under
isothermal compression, with the assumption that the pressure-volume
relation remains linear. Hence, in the optimization procedure we have
restricted the choice close to the equilibrium.

\begin{table}[h]
\begin{longtable}{|c||c|c||c|c|}
\hline 
&
\multicolumn{1}{c}{GGA}&
\multicolumn{1}{c||}{}&
\multicolumn{1}{c}{LDA}&
\multicolumn{1}{c|}{}\tabularnewline
\hline
&
$ZnCNi_{3}$&
$MgCNi_{3}$&
$ZnCNi_{3}$&
$MgCNi_{3}$\tabularnewline
\hline
\hline 
$a_{eq}$ (a.u.)&
7.2255&
7.3041&
7.0558&
7.1387\tabularnewline
\hline 
$B_{eq}$ (Mbar)&
0.3886&
0.3479&
0.4656&
0.4188\tabularnewline
\hline 
$B_{eq}^{'}$&
4.4106&
4.5255&
4.3444&
4.7813\tabularnewline
\hline
\end{longtable}

\caption{\label{tab-zncni3mg-eos}Comparison of the equation of state parameters
of cubic perovskite $ZnCNi_{3}$ with that of $MgCNi$ using the KKR-ASA
method as described in the text. }
\end{table}

Since the choice of the exchange correlation potential in the Kohn-Sham
Hamiltonian has proven to be sensitive in the structural characterization,
we have carried out the total energy minimization for two different
approximations, namely the LDA and GGA \cite{PRB-45-13244,prb-54-16533}.
The results are shown in Table.\ref{tab-zncni3mg-eos}. The GGA considerably
overestimates the lattice constant for either alloys, when compared
to the experimental values. For $MgCNi_{3}$, using the LDA description
of the exchange-correlation, the lattice constant was calculated as
$7.139$ a.u., with the bulk modulus and its pressure derivative as
$0.42$ Mbar and $4.78$ respectively. These values are consistent
with the earlier first-principles reports \cite{PRB-140507,PRB-100508}.
The underestimation in the lattice constant for $MgCNi_{3}$, however
when compared to the experiments, owe to the over-binding effects
in the LDA, and is a well known problem.

For $ZnCNi_{3}$ the equilibrium lattice constant calculated using
LDA yielded the value as $7.056$ a.u. which when compared to the
recent X-ray diffraction results \cite{sst-17-274} of $6.918$ a.u.,
was found to be an overestimation. However, the results of the present
calculations are consistent with the works of Johannes and Pickett
\cite{PRB-70-060507} who employed the FP-LAPW method. Note that the
consistency of the ASA calculations with that of the full-potential
counterparts owe to the inclusion of the muffin-tin correction \cite{PRL-55-600}.
The KKR-ASA calculations further finds the bulk modulus and its pressure
derivative of $ZnCNi_{3}$ as $0.46$ Mbar and $4.34$, respectively.
As mentioned above, the overestimation of lattice constant in LDA
is not so common, which suggests that the samples subjected to the
experiments may be sub-stoichiometric. This was also emphasized by
Johannes and Pickett \cite{PRB-70-060507} following the crystal structure
characterization of $MgC_{y}Ni_{3}$ alloys \cite{PhysicaC-1}. In
the latter, both experiments \cite{PhysicaC-1,PRB-024523} and theoretical
calculations \cite{PRB-72-064519} have shown that the lattice constant
decreases as the $C$ content in the material decreases.

\begin{figure}[h]
~

~

\includegraphics[%
  clip,
  scale=0.5]{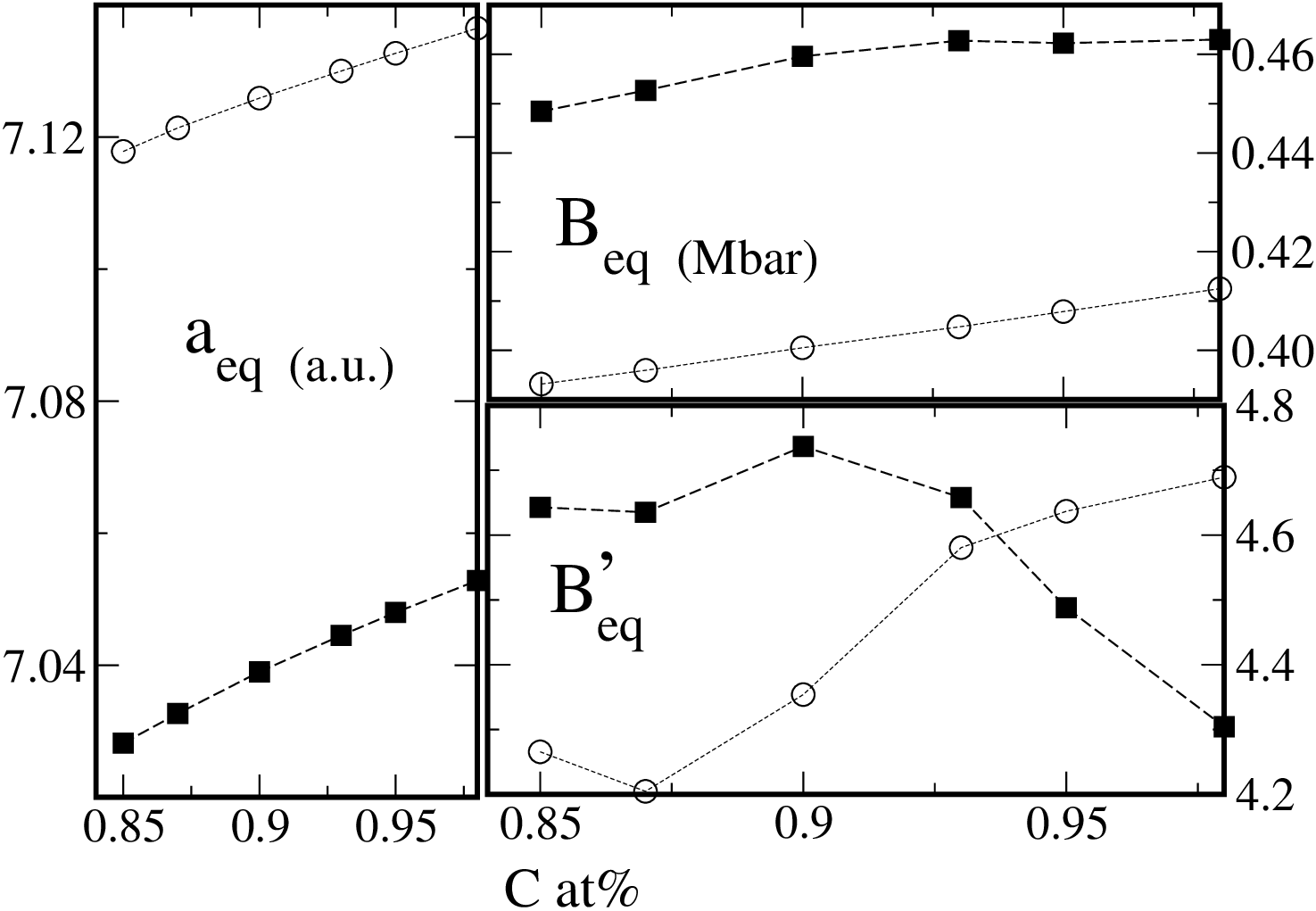}

\caption{\label{EqOfState}The variation in the equation of state parameters,
equilibrium lattice constant $a_{eq}$ ($a.u.$) , the bulk modulus
$B_{eq}$ (Mbar) and the pressure derivative of the bulk modulus,
as a function of $y$ in $MgC_{y}Ni_{3}$ (open circles) and $ZnC_{y}Ni_{3}$(filled
squares) calculated using the KKR-ASA-CPA method as described in the
text.}
\end{figure}

To look for the changes in the equation of state parameters as a function
of $C$ content in $MgC_{y}Ni_{3}$ and $ZnC_{y}Ni_{3}$ alloys, total
energy minimization was carried out. The variation is shown in Fig.\ref{EqOfState}.
For both $MgC_{y}Ni_{3}$ and $ZnC_{y}Ni_{3}$ alloys, the lattice
constant as well as the bulk modulus decrease as $C$ $at$\% decreases.
For $MgC_{y}Ni_{3}$ observed trend is consistent with the earlier
X-ray diffraction measurements. The rate of decrease in the lattice
constant is estimated as $0.142$ $a.u.$/$at$\%$C$ while for $ZnC_{y}Ni_{3}$,
the lattice constant was found to decrease at the rate of $0.189$
$a.u.$ per $at$\% of $C$. Though the lattice constant and bulk
modulus showed similar trend for either alloys, the change in the
pressure derivative of the bulk modulus as a function of $y$ characteristically
differed.

The pressure derivative of the bulk modulus measures the rate at which
the material becomes incompressible with increasing pressure, and
is sensitive to the softness of the equation of state. In the Debye
approximation for isotropic solids, which assumes a uniform dependence
of the lattice frequencies with volume, one may express the average
phonon frequency $\omega$ as $B_{eq}^{'}$$\propto$ $\,\frac{\delta\, ln\omega}{\delta\, lnV}$,
where $V$ is the equilibrium volume of the unit cell. Note that volume
for the vacancy-rich alloys decreases with decreasing $y$, while
$B_{eq}^{'}$ maps a different trend for $MgC_{y}Ni_{3}$ and $ZnC_{y}Ni_{3}$
alloys. Such a behaviour indicates that the properties associated
with the $MgC_{y}Ni_{3}$ lattice could be characteristically different
from that of the $ZnC_{y}Ni_{3}$ counterparts. Also, one may note
that the phonon spectrum for $MgCNi_{3}$ reveals that certain $C$
modes play a vital role in the materials superconducting properties
in addition to those of the $Ni$ modes \cite{PRB-092511,PRB-220504}.

\begin{figure}[h]
~

~

\includegraphics[%
  clip,
  scale=0.5]{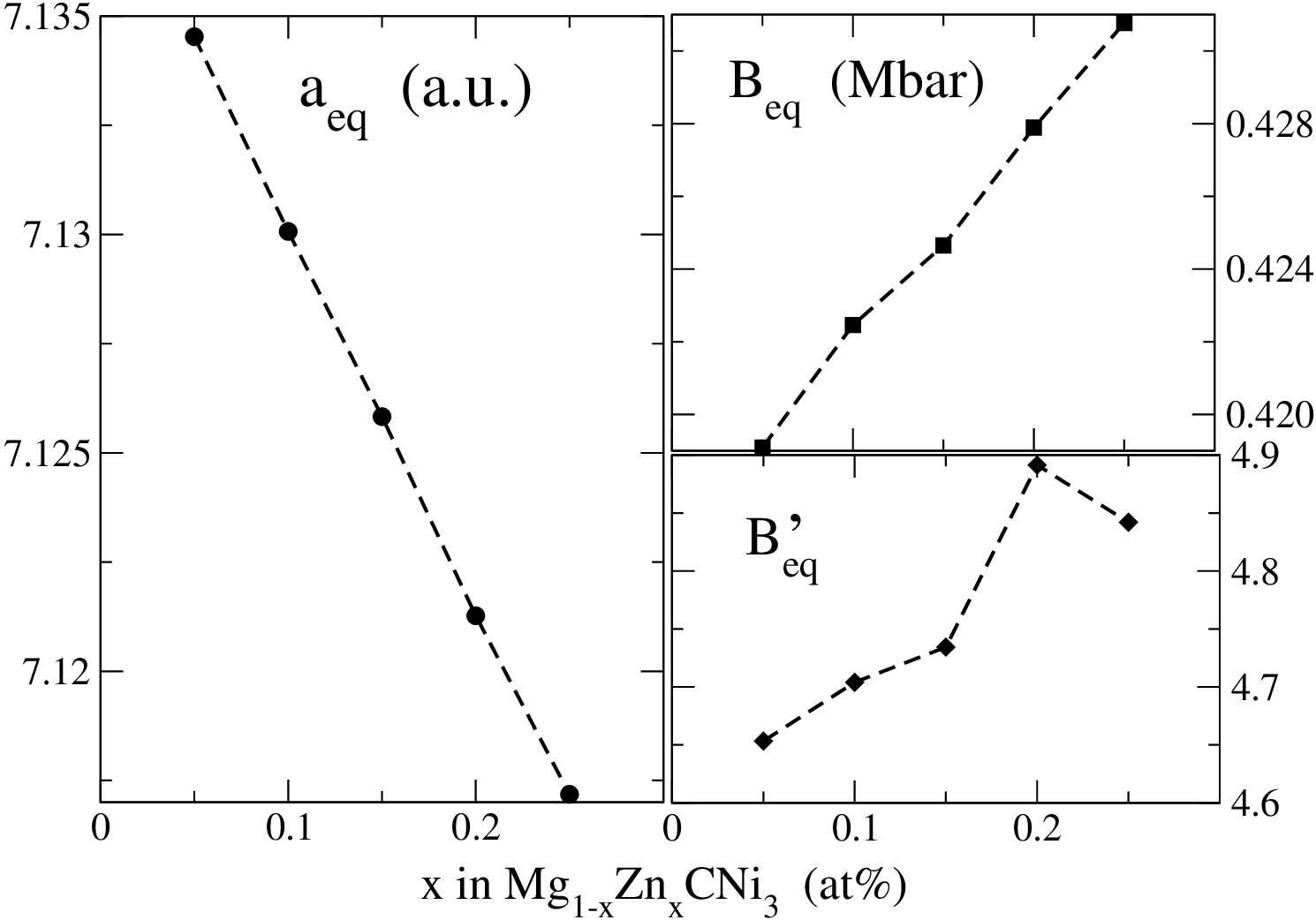}

\caption{\label{znxmg-eos}The variation in the equation of state parameters,
equilibrium lattice constant $a_{eq}$ ($a.u.$) , the bulk modulus
$B_{eq}$ (Mbar) and the pressure derivative of the bulk modulus,
as a function of $x$ in $Mg_{1-x}Zn_{x}CNi_{3}$ calculated using
the KKR-ASA-CPA method as described in the text.}
\end{figure}

Partial replacement of $Zn$ for $Mg$ in $MgCNi_{3}$ has shown that
the transition temperature decreases \cite{PhysicaC-160}. The findings
also conclude that the nature of pairing mechanism in $MgCNi_{3}$
is conventional \cite{PhysicaC-160}. To study the changes brought
about by $Zn$ substitutions in the $Mg$ sub-lattice of $MgCNi_{3}$,
we have carried out KKR-ASA-CPA calculations for $Mg_{1-x}Zn_{x}Ni_{3}$
alloys. In Fig.\ref{znxmg-eos}, we show the variation of the equation
of state parameters of $Mg_{1-x}Zn_{x}Ni_{3}$ alloys. The decrease
in the lattice constant is consistent with the previous experimental
report. The bulk modulus as well as its pressure derivative increases
as $x$ increases in $Mg_{1-x}Zn_{x}Ni_{3}$ alloys. This clearly
indicates that the average phonon frequency gets modulated as $Zn$
replaces $Mg$ in $MgCNi_{3}$.

\subsection{Electronic structure}

\begin{figure}[h]
~

~

\includegraphics[%
  clip,
  scale=0.5]{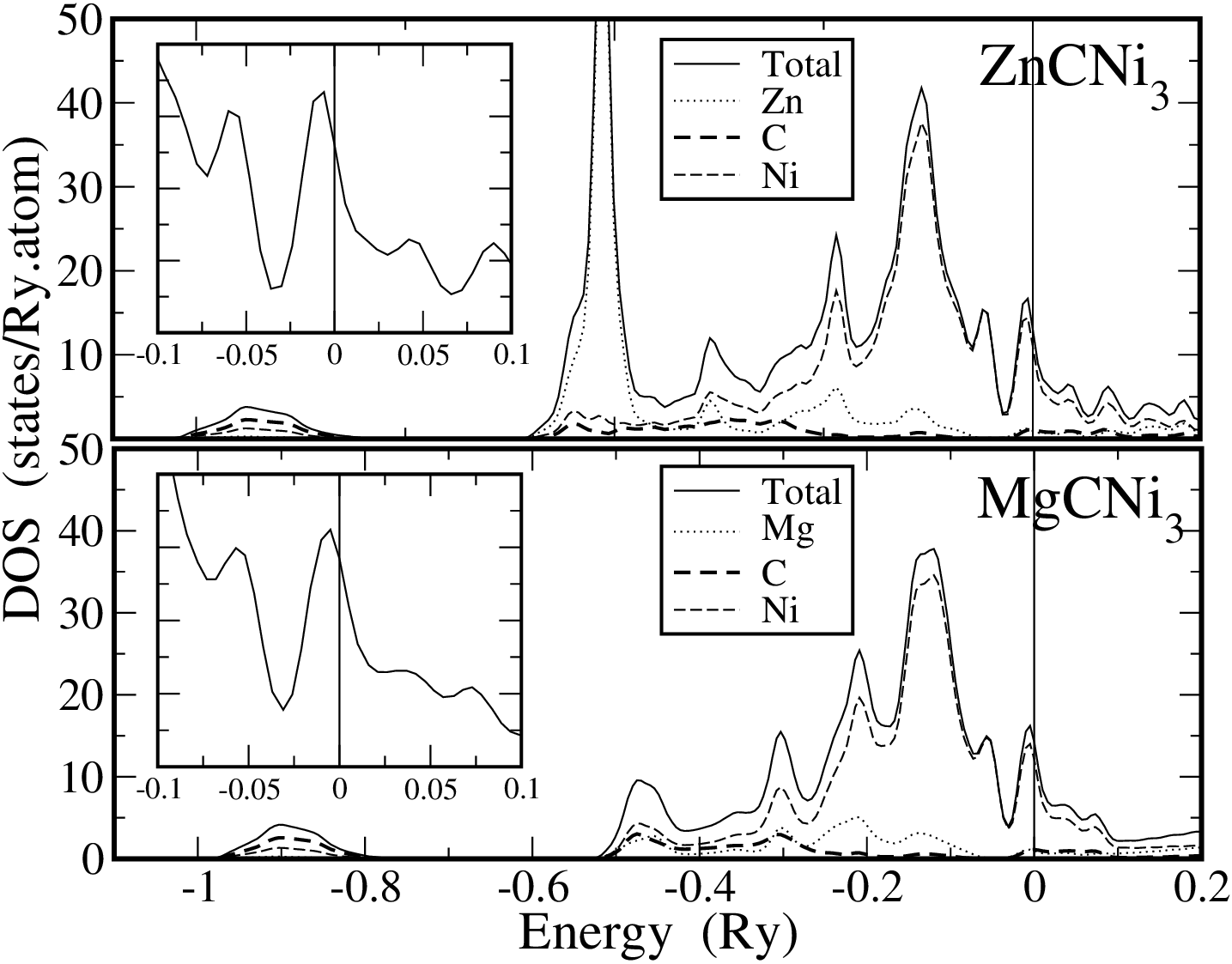}

\caption{\label{zncni3mg-Total-DOS}Comparison of the total and sub-lattice
resolved partial density of states of $ZnCNi_{3}$ and $MgCNi_{3}$,
calculated at their respective equilibrium lattice constants. The
vertical line through energy zero represents the alloy Fermi energy.
In the inset we show a blow up of the total density of states near
the Fermi energy.}
\end{figure}

In Fig.\ref{zncni3mg-Total-DOS}, we show the total and sub-lattice
resolved partial densities of states of $MgCNi_{3}$ and $ZnCNi_{3}$
calculated at their respective equilibrium lattice constants. The
characteristic features of both $MgCNi_{3}$ and $ZnCNi_{3}$ appear
more or less similar with an exception of a sharp peak in the energy
range $-0.6\leq E\leq-0.4$, characteristic of $Zn$ $3d$ states.
Being deep below on the energy scale compared to the Fermi energy,
which is zero on the scale shown, one may expect $Zn$ $d$ states
to be localized and thus behave atomically. While for $MgCNi_{3}$
a small peak characteristic of the $Mg$-$Ni$ bonding also appears
in this energy range, but is less pronounced. Furthermore, the states
near $E_{F}$ are predominantly $Ni$ $3d$ in character in both alloys,
with little admixture of the $C$ $2p$ states. One may also find
that the position of the $Ni$ $3d$ derived singularity is slightly
lower in the energy scale for $ZnCNi_{3}$ than for $MgCNi_{3}$,
which is consistent with the previous results. The $N(E_{F})$ and
the contributions to it from the sub-lattices are compared in Table.\ref{tab-1}. 

\begin{table}[h]
\begin{longtable}{|c|c|c|c|c|c|c|c|c|}
\hline 
&
$N(E_{F})$&
$Zn$($Mg$)&
$C$&
$Ni$&
$Ni$ $d_{xy(xz)}$&
$Ni$ $d_{yz}$&
$Ni$ $d_{x^{2}-y^{2}}$&
$Ni$ $d_{3z^{2}-1}$\tabularnewline
\hline
\hline 
$ZnCNi_{3}$&
13.005&
0.945&
1.076&
10.984&
4.002&
0.168&
1.296&
0.606\tabularnewline
\hline 
$MgCNi_{3}$&
14.557&
1.016&
1.199&
12.341&
4.509&
0.097&
1.474&
0.658\tabularnewline
\hline
\end{longtable}

\caption{\label{tab-1}Comparison of the total $N(E_{F})$ and sub-lattice
resolved density of states of $ZnCNi_{3}$ and $MgCNi_{3}$ expressed
in units of states/Ry atom. }
\end{table}

\begin{figure}[h]
~

~

\includegraphics[%
  clip,
  scale=0.5]{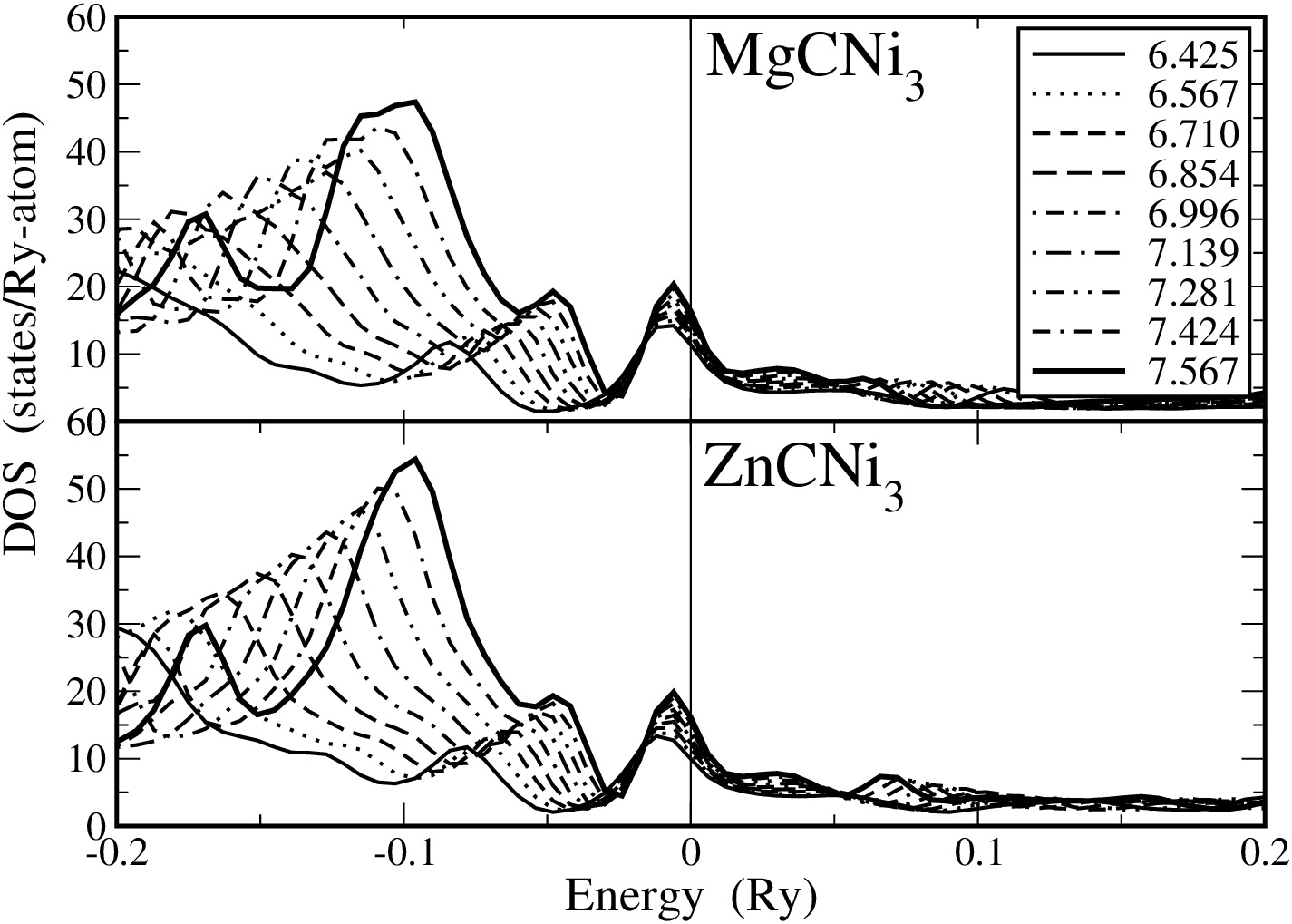}

\caption{\label{pressure-dos}Comparison of the change in the total density
of states near Fermi energy of $ZnCNi_{3}$ and $MgCNi_{3}$ for a
range of lattice constant as indicated in the figure. The vertical
line through energy zero represents the alloy Fermi energy. }
\end{figure}

The reported values of $N(E_{F})$ for $MgCNi_{3}$ are at variance
with the existing reports \cite{PRB-140507,PRL-027001,PRB-100508,JPCM-L595,PRB-66-024520,PRB-65-064525,PhysC-408-154,PhysB-337-95,jap-91-8504}.
It appears that the value is sensitive to the basic approximations
made in each type of the electronic structure method, and also to
the parameters like that of the choice of Wigner-Seitz radii, choice
of the exchange-correlation potential and others. However, under similar
approximations, it is clear that for $ZnCNi_{3}$ the $N(E_{F})$
reduces by $12$\% in comparison with $MgCNi_{3}$. This is consistent
with the earlier first-principles FP-LAPW calculations \cite{PRB-70-060507}.
The reduction in $N(E_{F})$ may be largely due to the smaller lattice
constant of $ZnCNi_{3}$, in comparison with $MgCNi_{3}$. The change
in the density of states, as well as in the $N(E_{F})$ as a function
of lattice constant is shown in Fig.\ref{pressure-dos}. Approximating
the variation of $N(E_{F})$ to be linear with respect to the lattice
constant, we find $dN(E_{F})$/$da$ to be $20.46$ and $22.02$ st/Ry
atom/a.u for $MgCNi_{3}$ and $ZnCNi_{3}$, respectively.

\begin{figure}[h]
~

~

\includegraphics[%
  clip,
  scale=0.5]{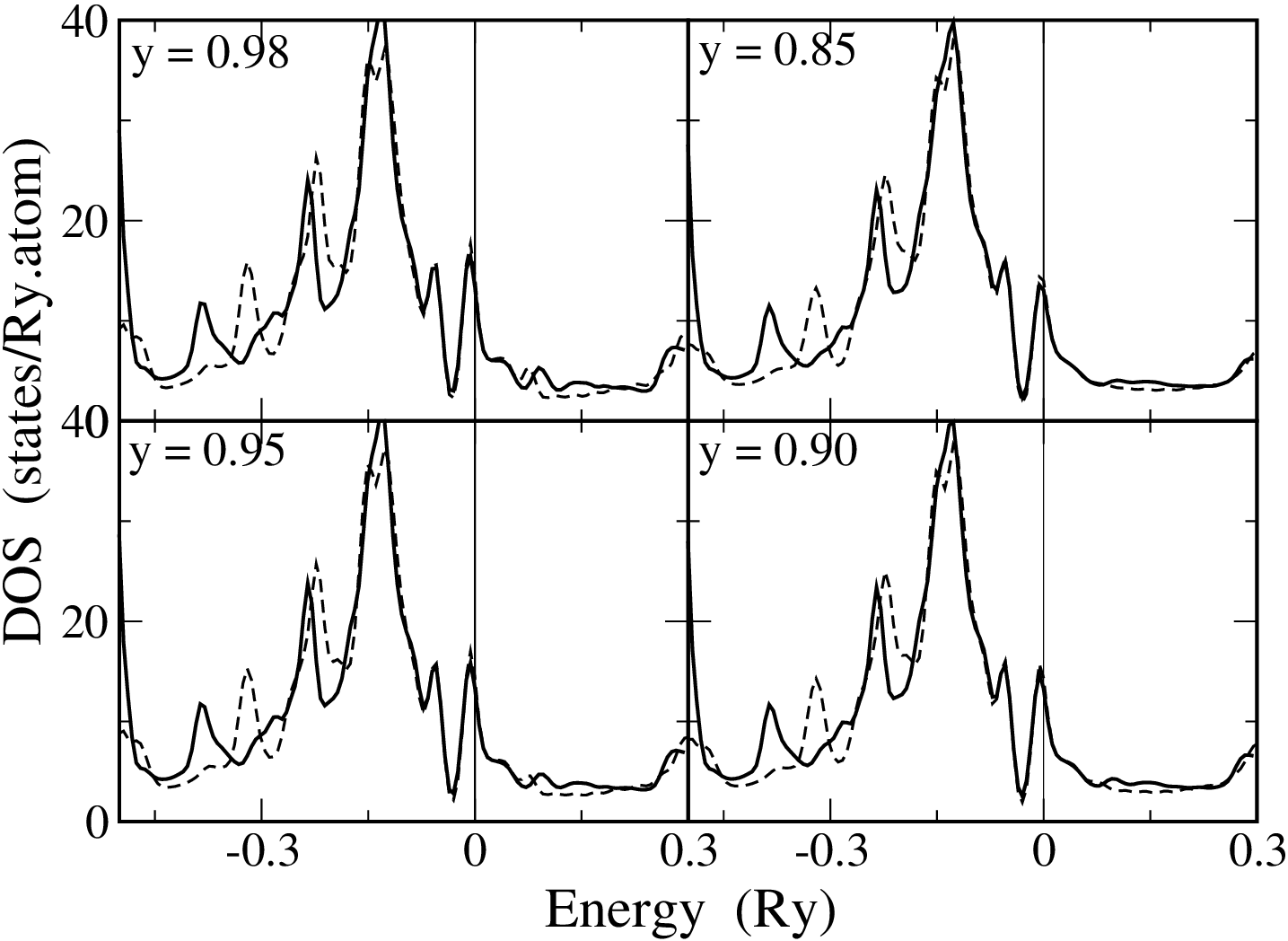}

\caption{\label{cxtotdos}Comparison of the change in the total density of
states of $ZnC_{y}Ni_{3}$ (solid line) and $MgC_{y}Ni_{3}$ (dashed
lines) alloys calculated at their equilibrium lattice constant with
$y$ as indicated. he vertical line through energy zero in each panel
show represents the Fermi energy.}
\end{figure}

\begin{figure}[h]
~

~

\includegraphics[%
  clip,
  scale=0.5]{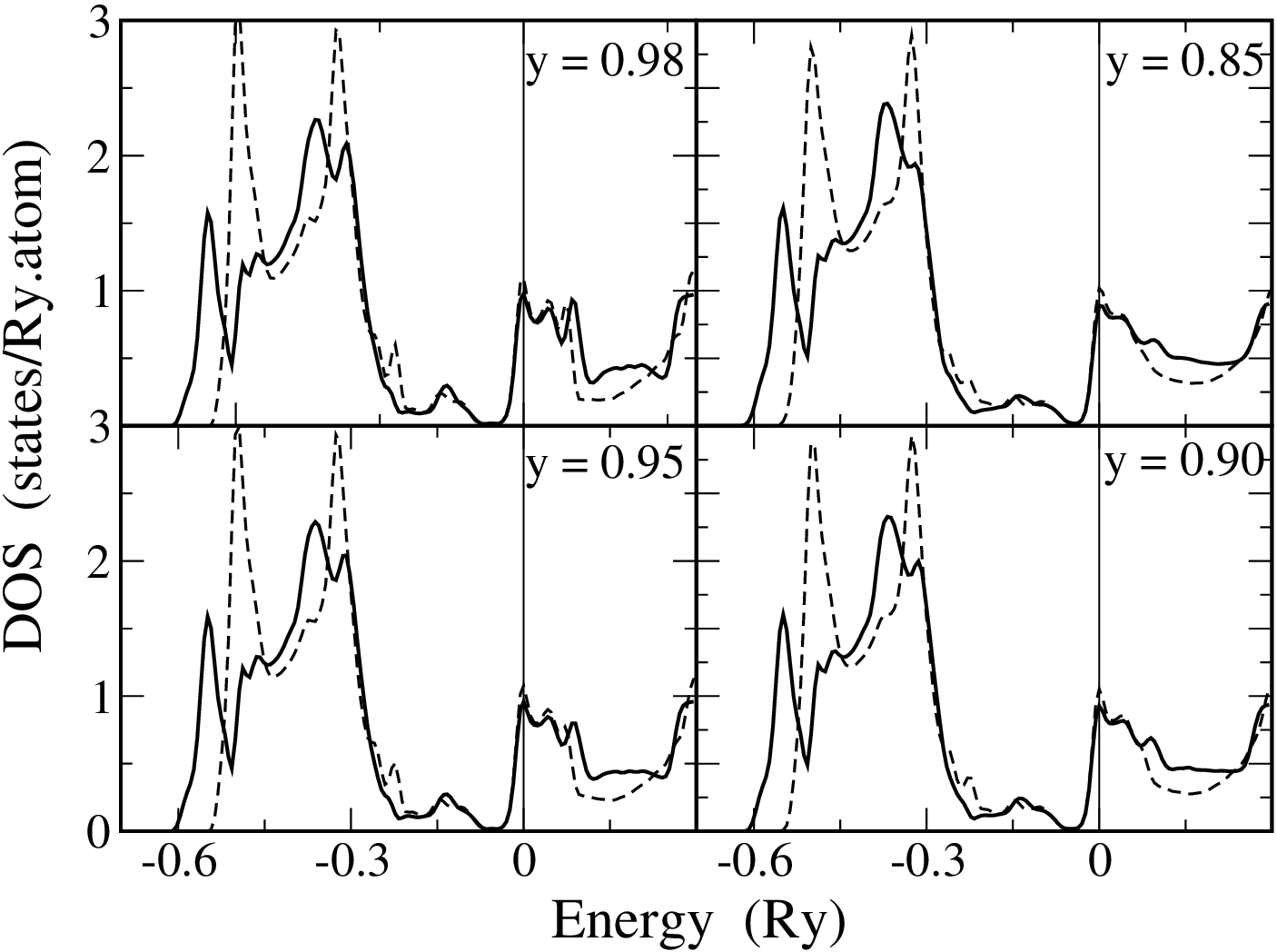}

\caption{\label{c2pdos}Comparison of the change in the sub-lattice resolved
$C$ $2p$ partial density of states of $ZnC_{y}Ni_{3}$ (solid line)
and $MgC_{y}Ni_{3}$ (dashed lines) alloys calculated at their equilibrium
lattice constant with $y$ as indicated. The vertical line through
energy zero in each panel represents the Fermi energy.}
\end{figure}

\begin{figure}[h]
~

~

\includegraphics[%
  clip,
  scale=0.5]{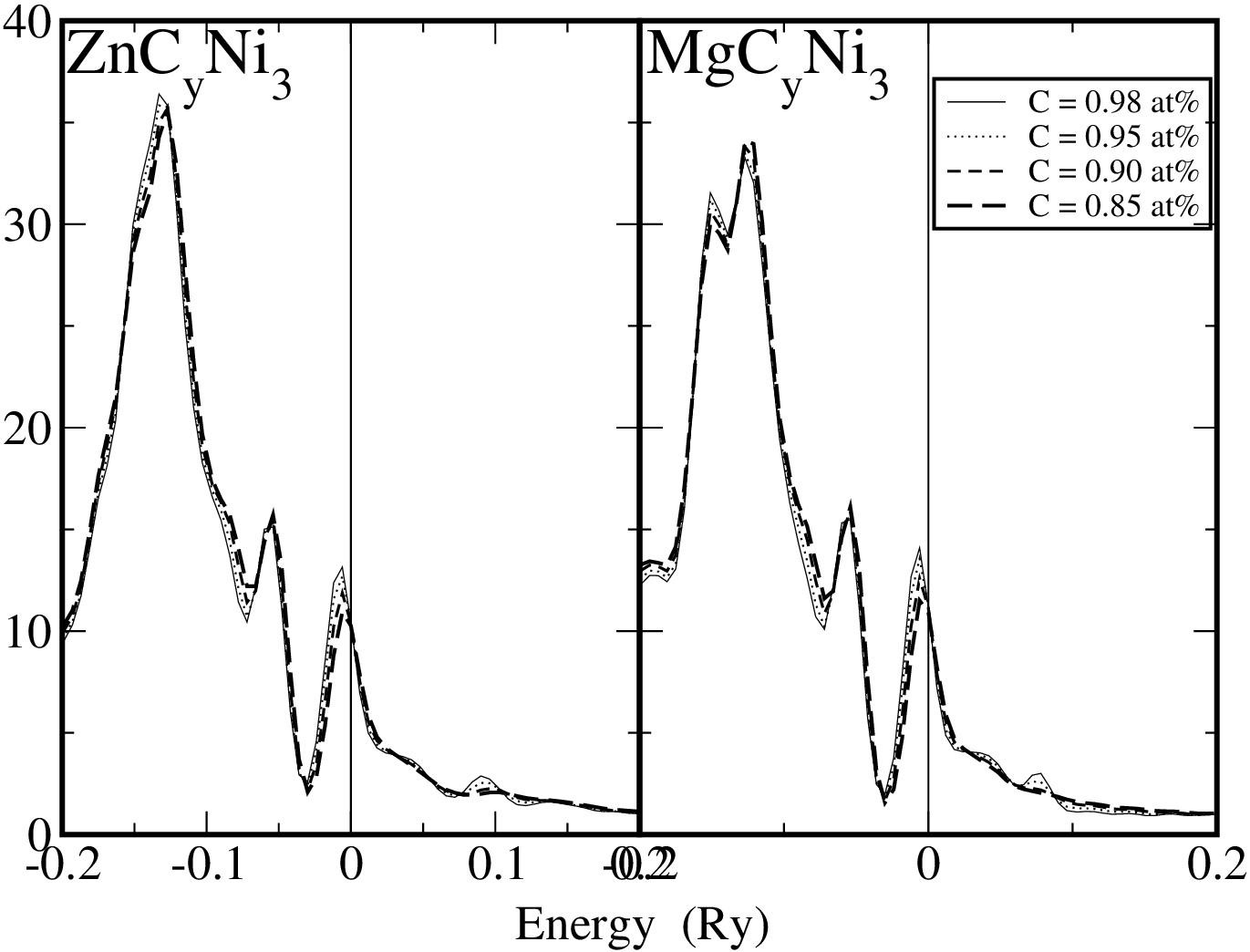}

\caption{\label{Ni3ddos}Comparison of the change in the sub-lattice resolved
$Ni$ $3d$ partial density of states of $ZnC_{y}Ni_{3}$ and $MgC_{y}Ni_{3}$
alloys, over a small energy window around Fermi energy, calculated
at their equilibrium lattice constant with $y$ as indicated. The
vertical line through energy zero in each panel represents the Fermi
energy.}
\end{figure}

\begin{figure}[h]
~

~

\includegraphics[%
  clip,
  scale=0.5]{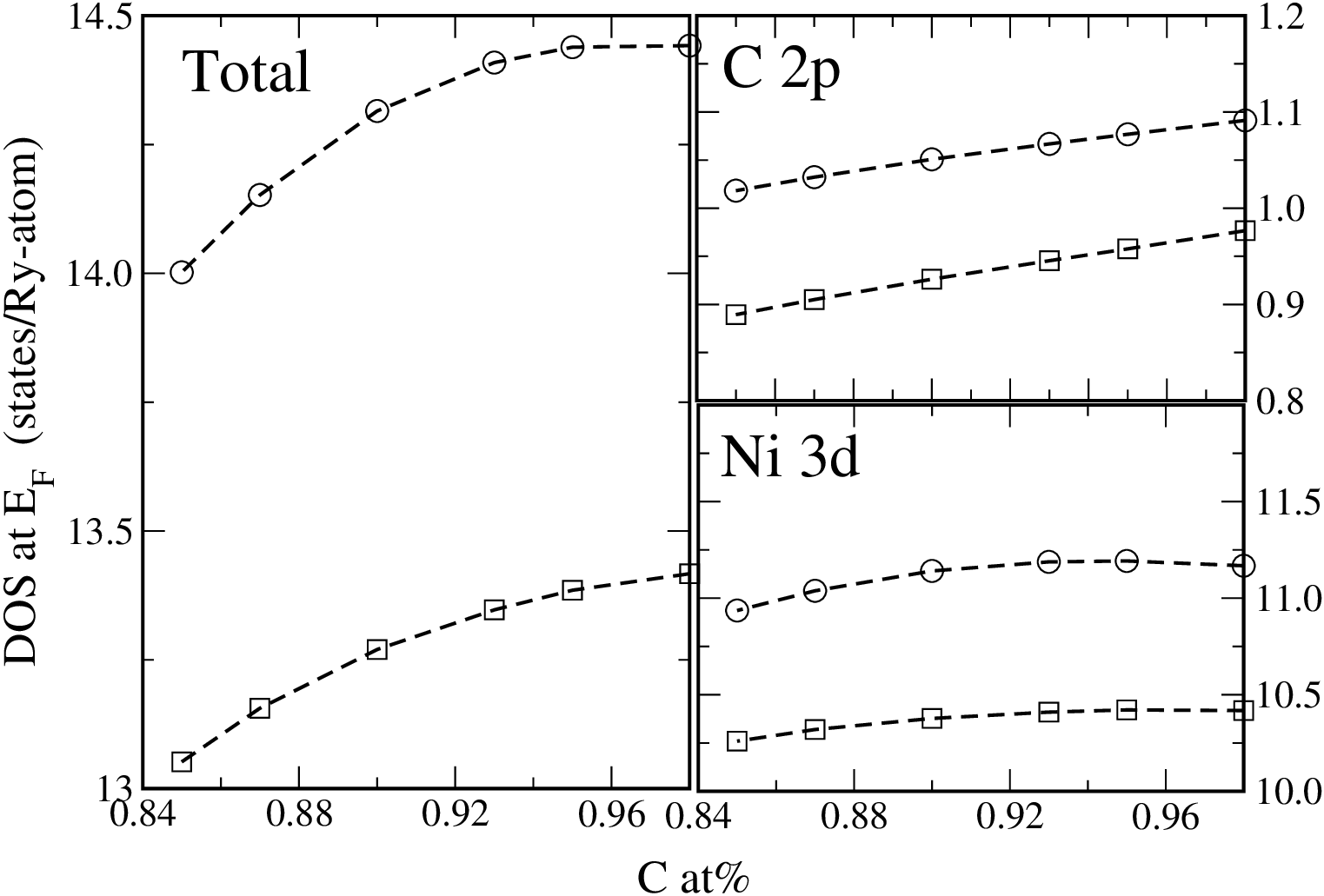}

\caption{\label{c2pni3ddos}Comparison of the change in the total, sub-lattice
resolved $C$ $2p$ and $Ni$ 3d partial densities of states at $E_{F}$
of $ZnC_{y}Ni_{3}$ (squares) and $MgC_{y}Ni_{3}$ (circles) as a
function of $y$. }
\end{figure}

To understand the changes in the electronic structure upon the introduction
of $C$ vacancies, we in Figs.\ref{cxtotdos}, \ref{c2pdos}, \ref{Ni3ddos}
and \ref{c2pni3ddos} show the changes in the total and sub-lattice
resolved $C$ $2p$ and $Ni$ $3d$ partial densities of states of
$ZnC_{y}Ni_{3}$ and $MgC_{y}Ni_{3}$ alloys calculated at their equilibrium
lattice constants. It follows from the figures that the change in
the distribution of states is more or less insignificant near the
Fermi energy, but states lower in energy undergo substantial changes. 

Upon creation of vacancies, a few of the $C$ 2$p$ - $Ni$ $3d$
bonds break, and result in charge redistribution. Note that the $CNi_{6}$
octahedral is a covalently built complex to which the cations at the
cube corners ($Zn$ and $Mg$) are thought to have donated their outermost
valence electrons. The crystal geometry suggests six $Ni$ atoms as
the first nearest neighbors to $C$ and eight $Mg$/$Zn$ atoms as
its second nearest neighbors. For $Ni$ the second nearest coordination
shell carries four $Mg$/$Zn$ atoms. The charge redistribution arising
due to the breaking of the $p$-$d$ bonds would be proportional to
the electro-positivity of the cation- $Mg$/$Zn$. Since $Mg$ is
more electro-positive than $Zn$, charge redistribution to the $Mg$/$Zn$
sub-lattices, as a function of vacancies would be more significant
in $MgCNi_{3}$ when compared to $ZnCNi_{3}$. This is consistent
with the fact that a larger fraction of the charge would be transferred
back to the $Mg$ sub-lattice, in $MgCNi_{3}$ in comparison with
that of the $Ni$ sub-lattice. 

\begin{figure}[h]
~

~

\includegraphics[%
  clip,
  scale=0.5]{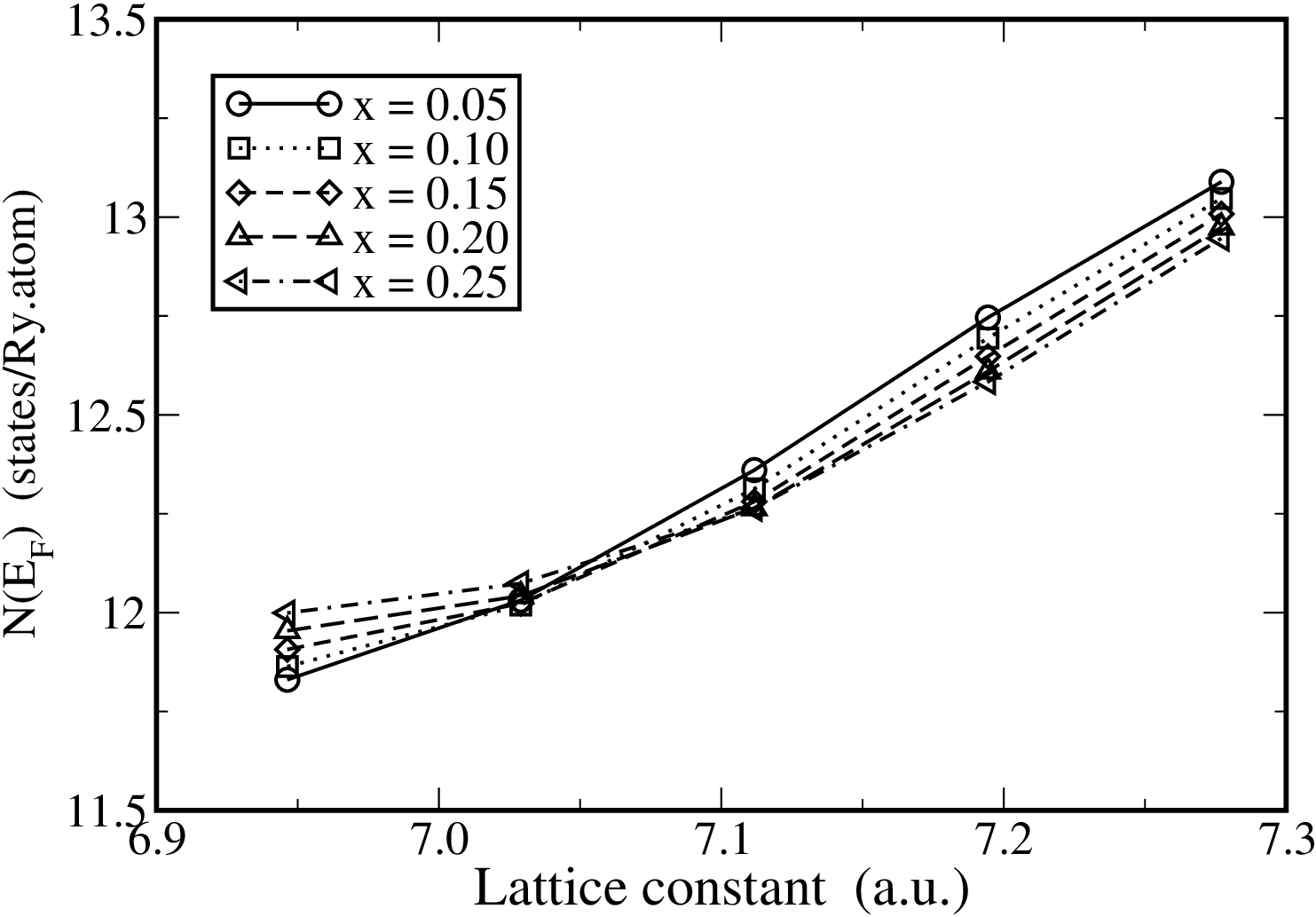}

\caption{\label{znxmg-latdos}The change in the total density of states at
the Fermi energy, $N(E_{F})$, in units of states/Ry atom as a function
of lattice constant in $Mg_{1-x}Zn_{x}CNi_{3}$ alloys. The values
of $x$ are as shown in the figure. }
\end{figure}

The change in the $N(E_{F})$ as a function of lattice constant in
$Mg_{1-x}Zn_{x}CNi_{3}$ alloys is shown in Fig.\ref{znxmg-latdos}.
One may find that $N(E_{F})$ decreases for all values of $x$, with
respect to lattice constant. However, $N(E_{F})$ as a function of
$x$, at the equilibrium lattice constant, was found to deviate a
little, as is evident from Fig.\ref{znxmg-latdos}. This clearly suggests
that the electronic structure properties are mainly governed by the
$CNi_{6}$ octahedra. The atoms occupying the cube corners i.e., $Mg$
and $Zn$, however, play a non-trivial role in determining the structural
properties.

\subsection{Hopfield parameter}

The Hopfield parameter $\eta$ has been regarded as a local ``chemical''
property of an atom in a crystal. It has been emphasized earlier that
the most significant single parameter in understanding the $T_{C}$
of a conventional superconductor is the Hopfield parameter \cite{douglass}.
For strong-coupling systems, the variation in $\eta$ is more important
than the variation of $\left\langle \omega^{2}\right\rangle $ in
changing $T_{C}$. Softening $\left\langle \omega^{2}\right\rangle $
often does enhance $T_{C}$, but a significant change in the magnitude
of $T_{C}$ depends largely on a significant change in the $\eta$
value rather than a small change in the corresponding $\left\langle \omega^{2}\right\rangle $.
As a matter of fact, we look for the changes in the $\eta$ from the
three sub-lattices of these perovskites as a function of lattice constant
as well as $y$ in $MgC_{y}Ni_{3}$ and $ZnC_{y}Ni_{3}$ alloys. Note
that for $MgCNi_{3}$, it has been reported that the superconducting
transition temperature $T_{C}$ increases upon application of external
pressure \cite{prb-68-092507,PRB-064510}. Besides, experiments remain
controversial on the strength of the electron-phonon interaction in
$MgCNi_{3}$ \cite{prb-67-052501,prb-70-174503,prb-68-092507,prb-67-094502}.
It has been suggested that $MgCNi_{3}$ may be a strongly-coupled
superconductor, however, the magnitude of $T_{C}$ being marginally
reduced due to the paramagnon interactions \cite{PRB-140507,prb-70-174503}.

\begin{figure}[h]
~

~

~

~

\includegraphics[%
  clip,
  scale=0.5]{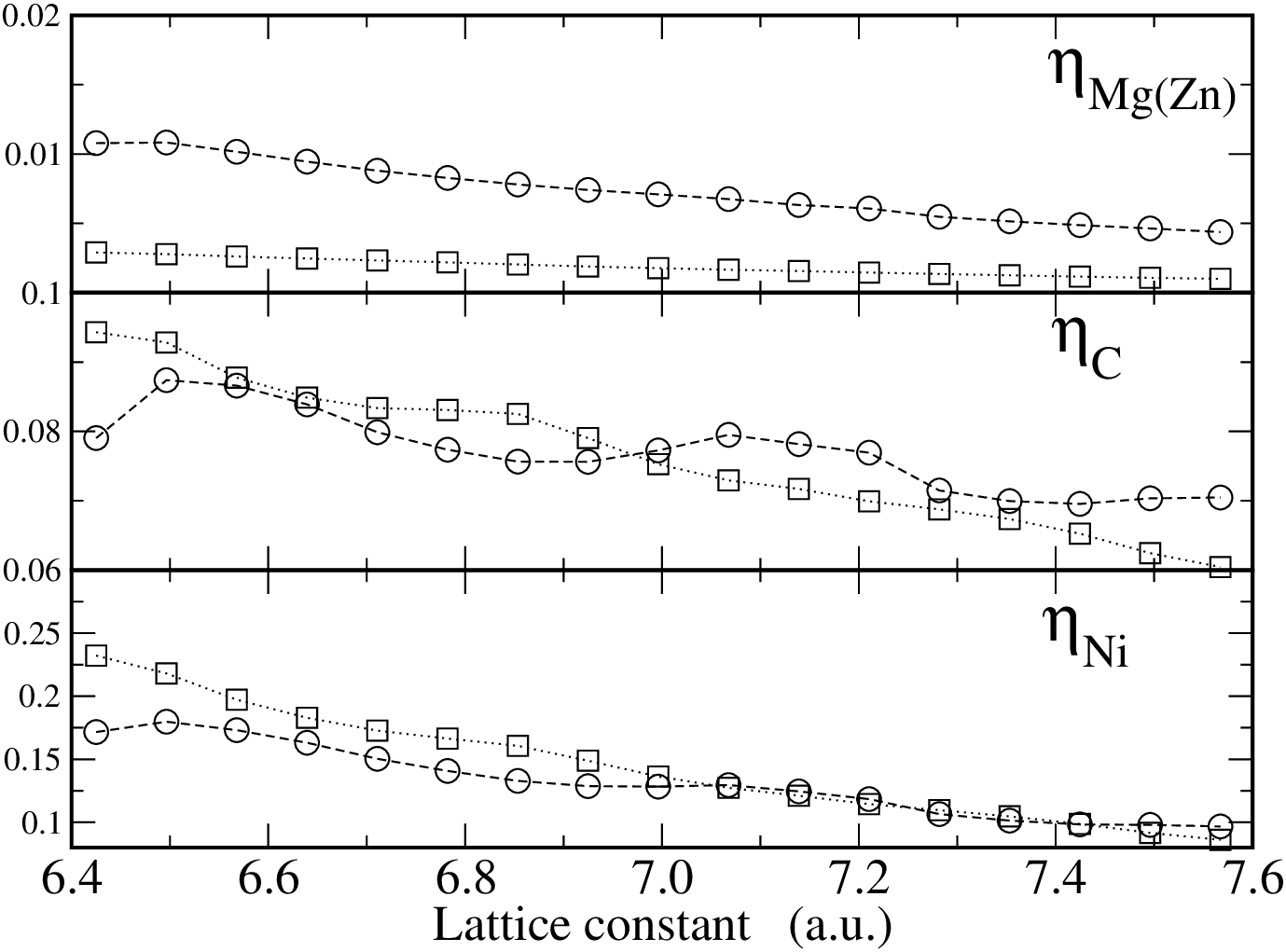}

\caption{\label{hopeta}Comparison of the change in the $\eta_{Mg/Zn}$, $\eta_{C}$
and $\eta_{Ni}$ as a function of lattice constant of $MgCNi_{3}$
(circles) and $ZnCNi_{3}$ (squares).}
\end{figure}

\begin{figure}[h]
~

~

\includegraphics[%
  clip,
  scale=0.5]{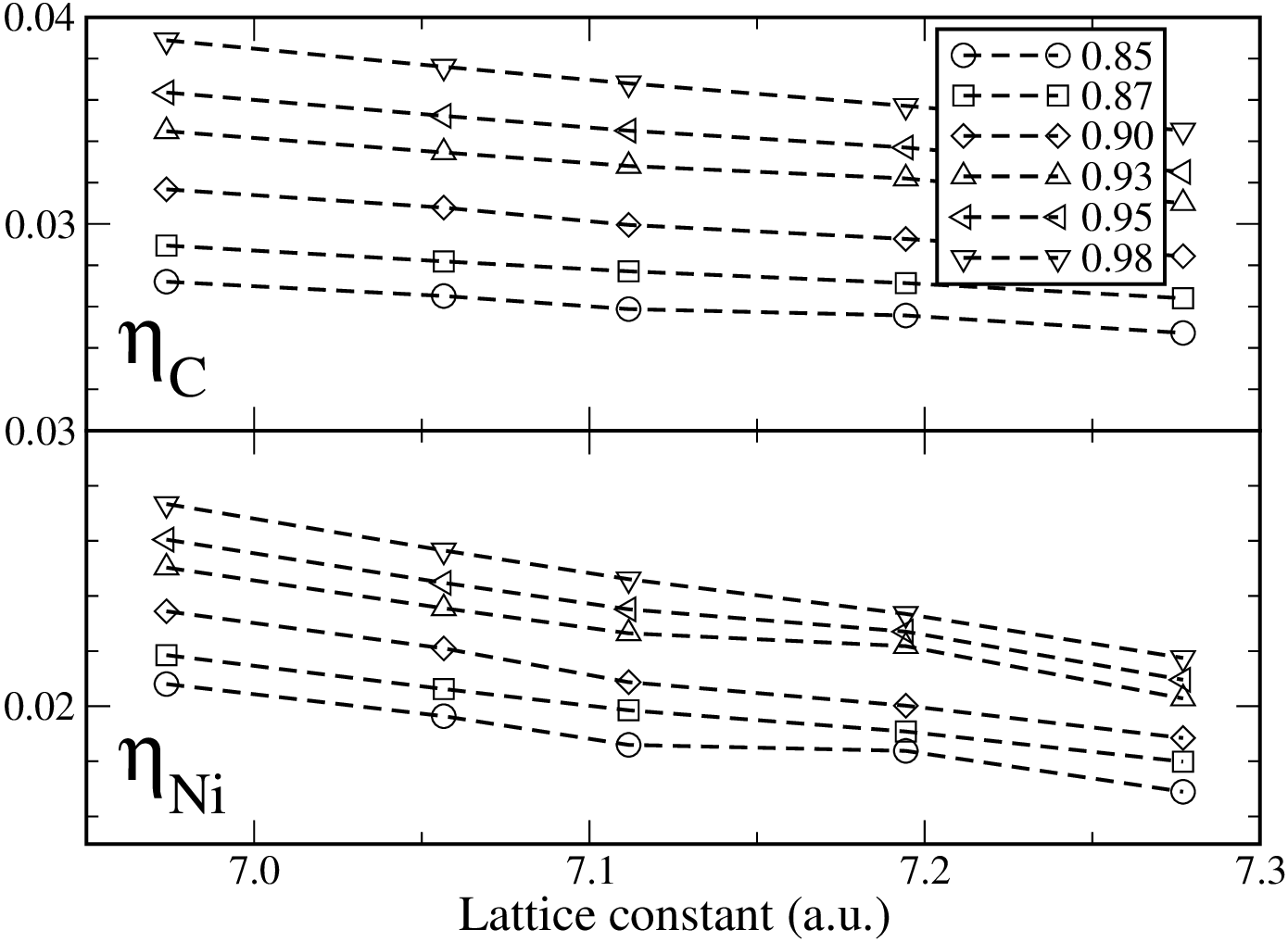}

\caption{\label{mgcxeta00}The change in the $\eta_{C}$ (upper panel) and
$\eta_{Ni}$ (lower panel) as a function of lattice constant in $MgC_{y}Ni_{3}$
alloys with $y$ as indicated.}
\end{figure}

\begin{figure}[h]
~

~

\includegraphics[%
  clip,
  scale=0.5]{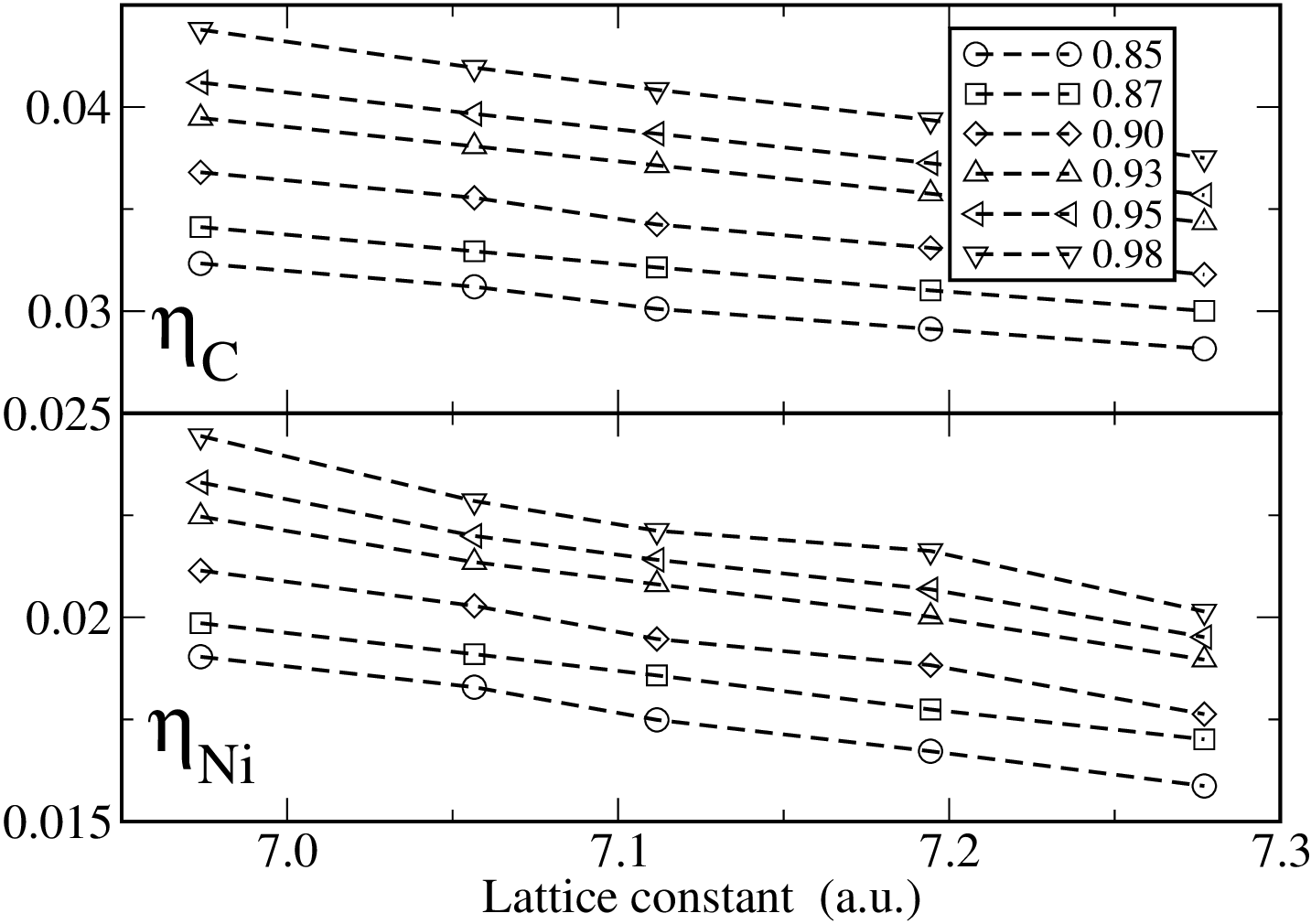}

\caption{\label{zncxeta00}The change in the $\eta_{C}$ (upper panel) and
$\eta_{Ni}$ (lower panel) as a function of lattice constant in $ZnC_{y}Ni_{3}$
alloys with $y$ as indicated.}
\end{figure}

In Fig.\ref{hopeta} we show the changes in the $\eta$ of $MgCNi_{3}$
and $ZnCNi_{3}$ as a function of lattice constant. It is clear from
Fig.\ref{hopeta} that the $\eta_{C}$ and $\eta_{Ni}$ linearly increase
as a function of decreasing volume in either alloys. If the change
in the average phonon frequency remains small, then either of these
alloys could enhance the transition temperature with respect to external
pressure. For $MgCNi_{3}$ this view is consistent with the previous
experimental results. Similar characteristic feature holds for the
vacancy-rich disordered alloys, the variation of which is shown in
Fig.\ref{mgcxeta00} and \ref{zncxeta00}

\begin{figure}[h]
~

~

\includegraphics[%
  clip,
  scale=0.5]{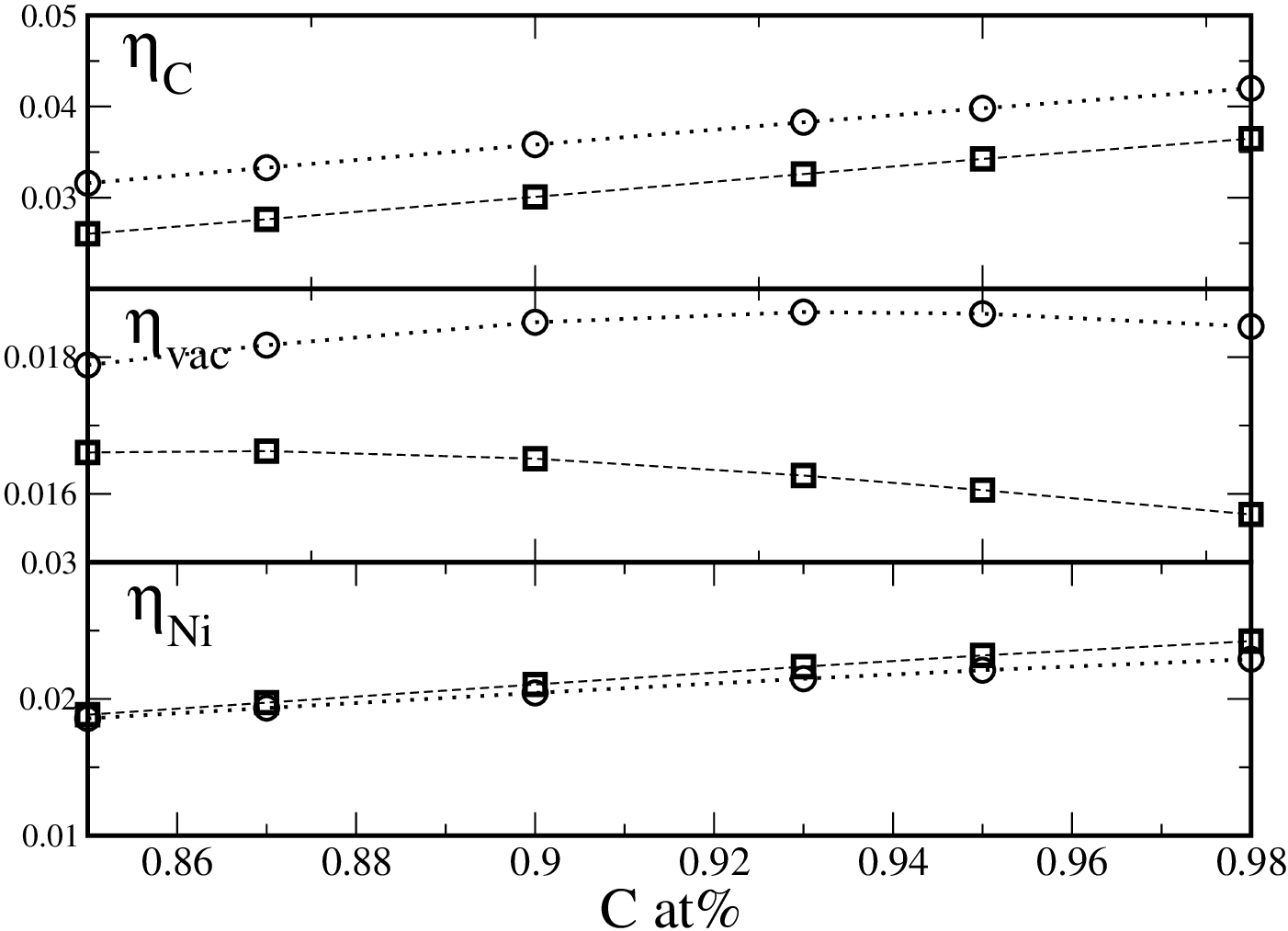}

\caption{\label{cxeta}Comparison of the change in the $\eta_{C}$, $\eta_{vac}$
and $\eta_{Ni}$ as a function of $y$ in $ZnC_{y}Ni_{3}$ (squares)
and $MgC_{y}Ni_{3}$ (circles) alloys. }
\end{figure}

To have an understanding in the variation of $\eta_{C}$, $\eta_{vac}$
and $\eta_{Ni}$ where $\eta_{vac}$ can be considered as the local
chemical property of the electrons in the empty sphere, we show in
Fig.\ref{cxeta} the change in these parameters as a function of $y$
in both $MgC_{y}Ni_{3}$ and $ZnC_{y}Ni_{3}$ alloys. One may find
that the variation of $\eta$ remains similar for both the alloys
as a function of decreasing $C$ content. 

\subsection{Magnetic properties}

Total energies from both the self-consistent, spin polarized and spin
unpolarized calculations remain degenerate for $MgCNi_{3}$ and $ZnCNi_{3}$
alloys at their equilibrium lattice constants. This unambiguously
shows that the materials are non-magnetic in nature. However, having
suggested that $MgCNi_{3}$ is on the verge of a ferromagnetic instability
\cite{PRB-140507,PRL-027001,prb-70-174503,prb-71-144516,PRL-257601},
and also that incipient magnetism in the form of spin-fluctuations
reside in the material, we attempt to compare the magnetic properties
of $MgCNi_{3}$ and $ZnCNi_{3}$ alloys using the fixed-spin moment
approach of alloy theory \cite{JPF-14-129}. 

\begin{figure}[h]
~

~

\includegraphics[%
  clip,
  scale=0.5]{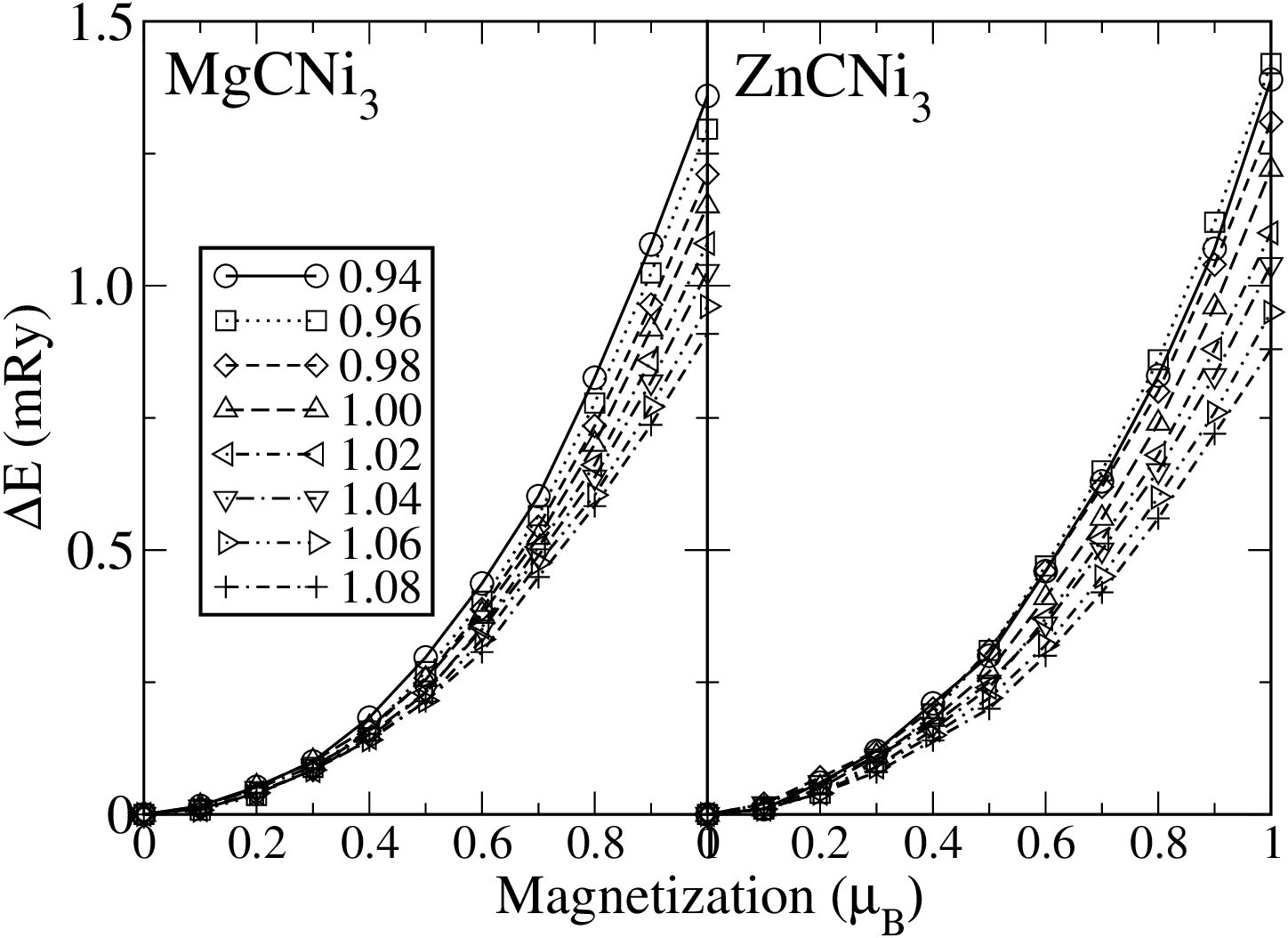}

\caption{\label{fsm-DE}Comparison of the change in the magnetic energy as
a function of magnetization in $MgCNi_{3}$ and $ZnCNi_{3}$ alloys
for a range of lattice constant ratio $a/a_{eq}$, where $a_{eq}$
is the equilibrium lattice constant of the respective alloys. }
\end{figure}

\begin{figure}[h]
~

~

\includegraphics[%
  clip,
  scale=0.5]{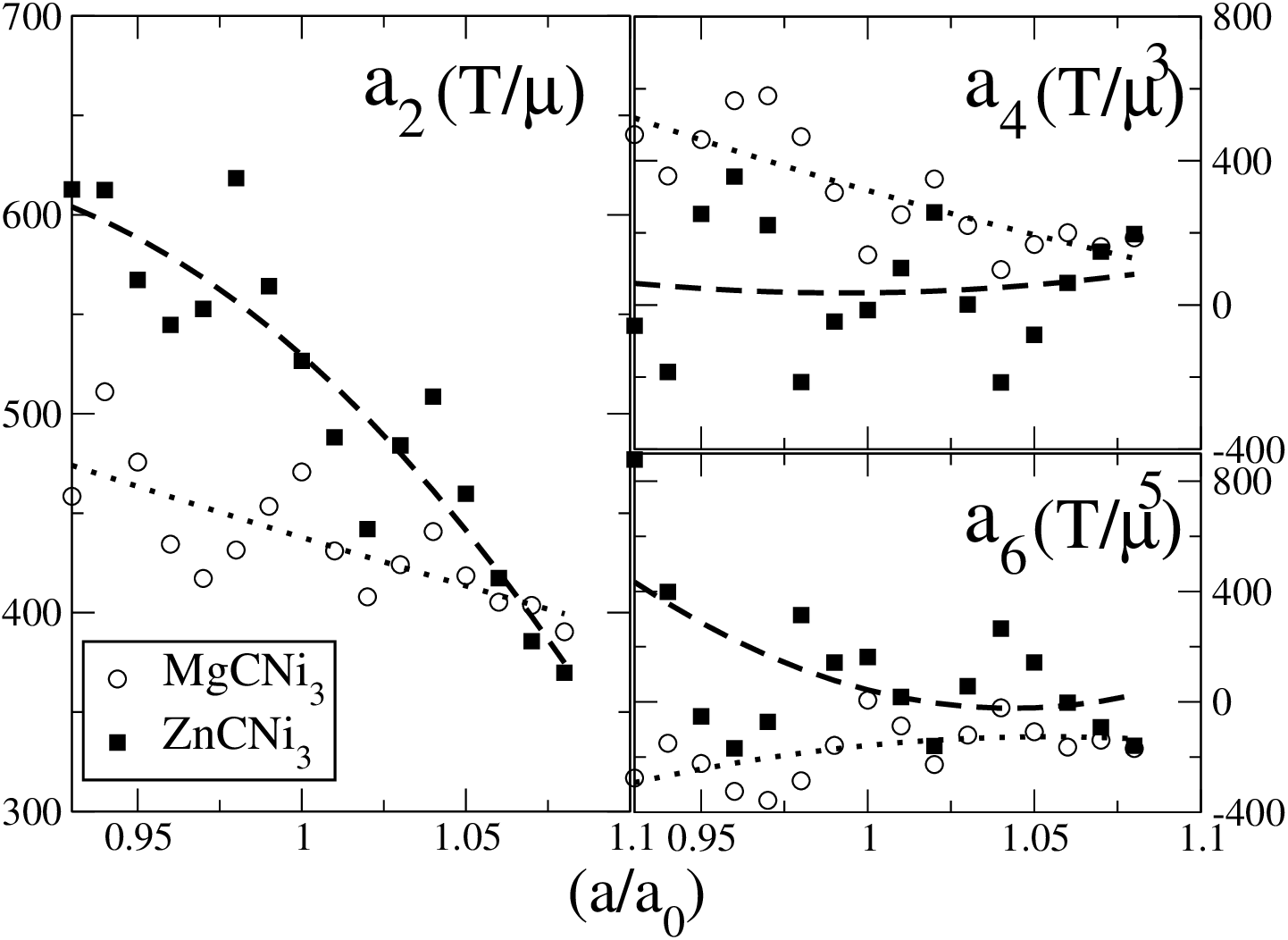}

\caption{\label{fsm-coeff}Comparison of the changes in the Landau coefficients
$a_{2}$, $a_{4}$ and $a_{6}$ as a function lattice constant in
$MgCNi_{3}$ and $ZnCNi_{3}$ alloys. The open circles and filled
squares are the calculated values and the best quadratic fit representing
these points are shown with dotted and dashed lines, respectively
for $MgCNi_{3}$ and $ZnCNi_{3}$ alloys. }
\end{figure}

Numerical calculations of magnetic energy $\Delta E(M)$ for $MgCNi_{3}$
and $ZnCNi_{3}$ are carried out at over a range of lattice constants.
The calculated results of $\Delta E(M)$ in the fixed-spin-moment
method are shown in Fig.\ref{fsm-DE}. The calculated $\Delta E(M)$
curves are fit to the form of a power series of $M^{2n}$ up to $n=3$,
for the polynomial as mentioned above. The variations of the coefficients,
$a_{2}$ in units of $\frac{T}{\mu_{B}}$, $a_{4}$ in $\frac{T}{\mu_{B}^{3}}$,
and $a_{6}$ in $\frac{T}{\mu_{B}^{5}}$ as a function of lattice
constant are shown in Fig.\ref{fsm-coeff}. The propensity of magnetism
can be inferred from the sign of the coefficient which is quadratic
in $M$, i.e., $a_{2}$. The coefficient $a_{2}$ is the measure of
the curvature and is positive definite when the total energy minimum
is at $M=0$, thus referring to a paramagnetic ground state. In general,
when $a_{2}$ becomes negative, it infers that there would exist a
minimum in the $\Delta E-M$ curve at a value other than $M=0$ referring
to a ferromagnetic phase at that value of $M$. The higher-order coefficients
$a_{4}$ and $a_{6}$ however are significant and they control the
variation of $\Delta E$ with respect to $M$. For example, for larger
values of $M$, $a_{4}$ and successively $a_{6}$ would dominate,
and if $a_{4}$($a_{6}$) tends to be negative it would show a dip
in the $\Delta E-M$ variation pointing towards a magnetic transition
at a higher value of $M$. This, in the first-principles characterization
of the magnetic properties of a material would refer to a possibility
of a metastable phase at relatively large values of external magnetic
fields. However, it has to be noted that calculations for large values
of $M$ can result in ambiguous results. Hence, it is suggested to
carry out calculations for smaller values of $M$ and use the above
mentioned polynomial function up to the minimum order, where the curve
fits with sufficient accuracy. 

Fig. \ref{fsm-coeff} shows that for smaller values of lattice constant,
the alloys show an enhanced paramagnetic character. One may also note
that the variation in $a_{4}$ and $a_{6}$ coefficients are oppositely
complimented and hence in the renormalized approach to include corrections
due to spin-fluctuations, as suggested by Yamada and Terao \cite{prb-59-9342},
they would cancel out in proportion preserving the trend in the variation
of $a_{2}$. Thus, it becomes likely that the incipient magnetic properties
associated with $MgCNi_{3}$ and $ZnCNi_{3}$ would decrease as a
function of decreasing lattice constant. 

\section{Conclusions}

First-principles syudy of the electronic properties of $MgCNi_{3}$
and $ZnCNi_{3}$, and also their non-stoichiometric alloys are carried
out using the density-functional-based KKR-ASA method. We find that
the lattice constant for $ZnCNi_{3}$ is overestimated, while for
$MgCNi_{3}$ it is underestimated. This suggests that the material
$ZnCNi_{3}$ subjected to experiments may be non-stoichiometric. As
a function of decreasing $C$ content in $MgC_{y}Ni_{3}$ and $ZnC_{y}Ni_{3}$
alloys, one finds an opposite trend in the variation of pressure derivative
of the bulk modulus, which is proportional to the averaged phonon
frequency. With electronic structure remaining essentially the same
for $MgC_{y}Ni_{3}$ and $ZnC_{y}Ni_{3}$, the results hint that non-stoichiometry
may have opposite effects. Note that for $0.9<y<1.0$, $MgC_{y}Ni_{3}$
alloys are feebly superconducting, while according to the conjecture
that has been made $ZnC_{y}Ni_{3}$ is not. It can thus be inferred
that the associated phonon modes in $ZnCNi_{3}$ and its disordered
alloys may be characteristically different when compared to the $MgCNi_{3}$
counterparts. A comparison of the phonon spectra of these alloys thus
become quite necessary to understand the absence of superconductivity
in $ZnCNi_{3}$, although it is iso-structural and iso-valent with
$MgCNi_{3}$.

\addcontentsline{toc}{chapter}{\numberline{}{Bibliography}}

\end{document}